\documentclass[aps,prc,preprint,groupedaddress,showpacs]{revtex4}
\usepackage{epsfig}
\usepackage{graphicx}
\usepackage{amsmath}
\usepackage{amsfonts,amssymb}
\usepackage{color}
  \usepackage{slashed}
  \usepackage{url}
  \usepackage{amsthm}
\usepackage{float}
\usepackage{bm}
\usepackage{verbatim}

\begin{document}

%\preprint{}

\title{Chiral symmetry in the low-energy limit of QCD at finite temperature}


\author{Marco Frasca}
\email[e-mail:]{marcofrasca@mclink.it}
%\thanks{}
\affiliation{Via Erasmo Gattamelata, 3 \\
             00176 Roma (Italy)}


\date{\today}

\begin{abstract}
We derive a non-local Nambu-Jona-Lasinio model from QCD with a form factor exactly obtained in the infrared limit. With this model, having all the parameters properly fixed through those of QCD, we consider the case of finite temperature and compute the solution of the gap equation at small temperature and small momenta and zero chemical potential. Taking the quark masses to be zero, it is proved that the theory undergoes a phase transition with a critical temperature exactly determined. These results prove unequivocally that the picture of the vacuum of QCD as a liquid of instantons is a very good approximation.
\end{abstract}

%\keywords{}

\maketitle


\section{Introduction}
%\label{}
%%%%%%%%%%%%

Currently, evidences for the existence of a phase transition in QCD,
% 5 November 2011 Asked by the referee
at finite temperature and chemical potential,
relies on lattice computations. This was firstly realized in a pioneering work by Fodor and Katz \cite{Fodor:2001pe} and further backed up by more recent studies as in \cite{Aoki:2006we} and \cite{Aoki:2009sc,Bazavov:2009zn} notwithstanding some criticisms were cast due to the infamous sign problem \cite{deForcrand:2010ys}. Studies on the behavior of QCD at finite temperature and density, from a theoretical standpoint, are generally performed using some phenomenological models like a Nambu-Jona-Lasinio model or a sigma model. The appearance of the infamous sign problem in lattice analysis has prompted some authors to introduce an imaginary chemical potential \cite{Alford:1998sd,de Forcrand:2002ci,D'Elia:2002gd} and the need for a consistent agreement between lattice computations and theoretical models has prompted the introduction of more general models using the Polyakov loop. In this framework some authors were able to prove the existence of a statistical confinement of quarks and give account of the phase diagram of QCD \cite{Fukushima:2008wg, Abuki:2008nm}. The relevant point to note for our aims is that, when quark masses are taken into account, there is not a real phase transition but rather a cross-over between a confined and a deconfined phase and, in a same range of temperatures, also a chiral broken symmetry is seen. But when the chemical potential and the masses of the quarks are taken to be zero, a first order phase transition is indeed expected at a given critical temperature.

It is clear from this situation that a proof based on first principles of at least a chiral breaking of symmetry, starting from the equations of QCD, does not exist yet. 
% 29 September 2011 Added refs on chiral perturbation theory
Efforts on this direction date back to 1980's where chiral perturbation theory come into play \cite{Gasser:1986vb,Gasser:1987ah,Rajagopal:1992qz} but did not produce a value for the critical temperature. 
%
%This 
The difficulty relies essentially on our impossibility to obtain a model for the low-energy behavior of QCD directly from the theory. Quite recently, we were able to prove that a non-local Nambu-Jona-Lasinio model is indeed such a low-energy limit for QCD \cite{Frasca:2008zp,Frasca:2008tg,Frasca:2010iv} and this result has been also obtained by Kei-Ichi Kondo \cite{Kondo:2010ts}. The crucial point in our derivation has been an analytical closed form for the gluon propagator in the limit of very low energies \cite{Frasca:2007uz,Frasca:2009yp}.

The form of the gluon propagator is an essential cornerstone result that permits to perform explicitly a lot of low-energy computations directly from the equations of QCD. This is clearly shown by a recent paper by Hell et al. \cite{Hell:2008cc}. These authors were able to give a complete account of a non-local Nambu-Jona-Lasinio model, both at zero and finite temperature, but the form of the propagator was only guessed through the idea that the ground state of QCD is that of a liquid of instantons \cite{Schafer:1996wv}. We will see below how successful is such a guess as our scenario is perfectly consistent with this view. The point is that now,
% 
% 28 September 2011 Removed by referee's request
%having completely determined from QCD the gluon propagator in the low-energy limit, 
%
we 
% 5 November 2011 referee's complaints
%are able once and for all to prove
will see
that the theory, at zero chemical potential and zero quark masses, indeed undergoes a symmetry breaking at low temperature and we are in a position to obtain the critical temperature just computed on the lattice. We just note that the value of the critical temperature obtained from lattice computation is yet a debated matter as two competing groups obtain not exactly the same value. But for our aims it is enough to be in the right range.

This paper is so structured. In sec.\ref{sec1} we present the results on the infrared limit of QCD. In sec.\ref{sec2}we show again, to make the paper self-contained, a derivation of the non-local Nambu-Jona-Lasinio model from QCD. In sec.\ref{sec3} we solve the gap equation at low temperature and low momenta giving the main result. In sec.\ref{sec4} we give the conclusions. 

%%%%%%%%%%%%
\section{QCD in the infrared limit}
\label{sec1}
%%%%%%%%%%%%

As usual, our starting point will be the generating functional of QCD. We take
\begin{eqnarray}
   S_{QCD}&=&-\frac{1}{4}\int d^4x{\rm Tr}F^2+\int d^4x\sum_q\bar q(x)\left(i\slashed{\partial}-g\frac{\lambda^a}{2}\slashed{A}^a\right)q(x) \nonumber \\
   &&-\int d^4x(\bar c^a\partial_\mu\partial^\mu c^a+g\bar c^a f^{abc}\partial_\mu A^{b\mu}c^c)
\end{eqnarray}
being $F^a_{\mu\nu}=\partial_\mu A^a_\nu-\partial_\nu A_\mu^a+gf^{abc}A^b_\mu A^c_\nu$, $g$ the coupling that in this case is dimensionless, $q(x)$ are the quark fields, $A_\mu^a(x)$ the vector potentials of the Yang-Mills field and $c^a$ the ghost field. So, it straightforward to write down
\begin{eqnarray}
%
% 28 September 2011 Corrected N to \cal N after referee's request
%
   Z_{QCD}[j,\bar\eta,\eta,\bar\epsilon,\epsilon]&=&{\cal N}\int[dA][d\bar q][dq][d\bar c][dc]
   e^{iS_{QCD}}e^{i\int d^4x\sum_q[\bar\eta_q(x)q(x)+\bar q(x)\eta_q(x)]}\times \nonumber \\
   &&e^{i\int d^4xj_\mu^a(x)A^{\mu a}(x)}
   e^{i\int d^4x(\bar\epsilon^a c^a+\bar c^a\epsilon^a)}.
\end{eqnarray}
Our aim is to find a proper approximation in the low-energy limit. We will perform an expansion in the inverse of the 't Hooft coupling. In order to manage this functional, it appears essential to find a way to reduce this theory to a simpler one. What could make the theory manageable is to find a set of classical solutions, in the proper infrared limit of the coupling going to infinity, to start a perturbation series for a quantum field theory that holds in the same approximation of a strong coupling.
% 29 September 2011
A posteriori, we will verify the soundness of our choice of the classical solutions to build up a quantum field theory by comparison with numerical solutions on the lattice and for Dyson-Schwinger equations. 

% 29 September 2011 Modified due to referee's request

\subsection{Gluon propagator}

%So, 
With this aim in mind, we have recently proved the following theorem, holding just for {\em classical solutions} and producing an asymptotic mapping between the scalar field and the Yang-Mills theory in the limit of the coupling going to infinity:

\newtheorem*{thm}{Mapping Theorem}
\begin{thm}
An extremum of the action
\begin{equation}
\nonumber
    S = \int d^4x\left[\frac{1}{2}(\partial\phi)^2-\frac{\lambda}{4}\phi^4\right]
\end{equation}
is also an extremum of the SU(N) Yang-Mills Lagrangian when one properly chooses $A_\mu^a$ with some components being zero and all others being equal, and $\lambda=Ng^2$, being $g$ the coupling constant of the Yang-Mills field, when only time dependence is retained. In the most general case the following mapping holds
\begin{equation}
\nonumber
    A_\mu^a(x)=\eta_\mu^a\phi(x)+O(1/\sqrt{N}g)
\end{equation}
being $\eta_\mu^a$ constant, that becomes exact for the Lorenz gauge.
\end{thm}

A first proof of this theorem was given in \cite{Frasca:2007uz} and, after a criticism by Terence Tao, a final proof was presented in \cite{Frasca:2009yp} also agreed with Tao \cite{tao}. In the following we give a sketchy proof for the sake of completeness, 
% 29 September 2011 Added on referee's request
but it should be kept in mind that here we are working yet with classical solutions
. So, let us consider the equation of motion of the scalar field
\begin{equation}
\label{eq:phi3}
   \partial^2\phi+\lambda\phi^3=0.
\end{equation}
Now, we consider a gradient expansion for this equation in the following way. Let us rescale the time variable as $t\rightarrow\sqrt{\lambda}t$. The above equation becomes
\begin{equation}
   \partial_t^2\phi+\phi^3=\frac{1}{\lambda}\triangle\phi
\end{equation}
and we are in a position to do perturbation theory on this equation in the limit $\lambda\rightarrow\infty$ setting $\phi=\sum_{n=0}^\infty\lambda^{-n}\phi_n$. 
% 5 November 2011 Added for a referee's request of clarification
We note at this point a peculiarity of perturbation expansions for nonlinear differential equations. Let us consider the small perturbation case and just rescale the field as $\phi\rightarrow\xi\phi$ being $\xi$ a function of the coupling $\lambda$. Applying this rescaling to eq.(\ref{eq:phi3}) we get
\begin{equation}
   \partial^2\phi+\lambda\xi^2\phi^3=0.
\end{equation}
Now we take $\lambda'=\lambda\xi^2$ and our perturbation expansion is now for $\lambda'$. So, our weak perturbation series seems just arbitrary and the same happens to the strong perturbation series we are considering, the rescaling in time conspiring to this. Indeed, our strong perturbation series is just dual to the weak one and must share the same properties (see \cite{Frasca:1998ch}). What makes inessential this arbitrariness is that the perturbation series must be mathematically consistent and one must impose $\lambda\rightarrow 0$ for a weak perturbation and $\lambda\rightarrow\infty$ for the strong one. But these are formally an infinitesimal quantity and an infinite one and multiplying them by a constant is inessential: Our expansion will just get an overall multiplying constant $\xi$.
So, at the leading order equation, one has the equation $\partial_t^2\phi_0+\phi_0^3=0$ that admits the solution
\begin{equation}
   \phi_0=\mu 2^\frac{1}{4}{\rm sn}\left(\frac{1}{2^\frac{1}{4}}\mu t+\theta\right)
\end{equation}
being $\mu$ and $\theta$ two integration constants that can be taken depending on space variables. But if these are taken exactly constant we have discovered a set of exact solutions of the equation we started from. Better, now one can do a Lorentz boost and transform this in a covariant set of massive exact solutions \cite{Frasca:2009nt}. In this case, doing a Lorentz boost boils down to a resummation of all the perturbation series in the inverse of $\lambda$. 

Now, let us consider Yang-Mills equations for a generic SU(N) group and a generic gauge:
\begin{equation}
\partial^\mu\partial_\mu A^a_\nu-\left(1-\frac{1}{\alpha}\right)\partial_\nu(\partial^\mu A^a_\mu)+gf^{abc}A^{b\mu}(\partial_\mu A^c_\nu-\partial_\nu A^c_\mu)+gf^{abc}\partial^\mu(A^b_\mu A^c_\nu)+g^2f^{abc}f^{cde}A^{b\mu}A^d_\mu A^e_\nu = 0.
\end{equation}
As for the scalar field, we implement a strong coupling expansion with the rescaling $t\rightarrow \sqrt{N}gt$ and imposing the expansion $A_\mu^a=\sum_{n=0}^\infty(\sqrt{N}g)^{-n}A_\mu^{a(n)}$. At the leading order of the expansion one get the equations
\begin{equation}
   \partial_t^2A^{a(0)}_\nu-\left(1-\frac{1}{\alpha}\right)\partial_t^2 A^{a(0)}_0\delta_{\nu 0}+
   f^{abc}f^{cde}A^{b\mu(0)}A^{d(0)}_\mu A^{e(0)}_\nu = 0.
\end{equation}
Now, one can always find a set of components of the Yang-Mills field, chosen to be all equal, that reduces this leading order equation to the one of the scalar field $\partial_t^2\phi_0+\phi_0^3=0$ provided we do the identification $\lambda=Ng^2$. We can write
\begin{equation}
   A_\mu^a(t,0)=\eta_\mu^a\phi(t,0)+O\left(\frac{1}{\sqrt{N}g}\right).
\end{equation}
A Lorentz boost restores covariance and makes $\eta$ coefficients depending on momenta. The $\eta$ coefficients can be fixed through a gauge choice. E.g. in the Landau gauge one can have
\begin{equation}
\label{eq:etaeta}
   \eta_\mu^a\eta_\nu^b=\delta_{ab}\left(\eta_{\mu\nu}-\frac{p_\mu p_\nu}{p^2}\right).
\end{equation}
We do a couple of important considerations. Firstly, we see that gauge invariance is not lost due to the mapping and this mapping is an asymptotic one holding in the limit of the coupling going to infinity. So, these classical solutions can be used to work in the infrared limit in a quantum field theory preserving the substantial physical behavior in the ultraviolet limit in the quantum theory for the scalar and the Yang-Mills fields (triviality and asymptotic freedom respectively).

%Applying the above mapping theorem, holding in the large coupling limit, QCD generating functional becomes:
%\begin{eqnarray}
%Z[\eta,\bar\eta,j]=\int \prod_q[dq][d\bar q][d\phi]
%e^{i(N^2-1)\int d^4x\left[\frac{1}{2}(\partial\phi)^2-\frac{Ng^2}{4}\phi^4\right]}\times&& \\
%e^{i\int d^4x\sum_q\bar q(x)\left[\gamma\cdot\left(i\partial-g\frac{\lambda\cdot\eta}{2}\phi\right)-m_q\right]q(x)}
%e^{i\int d^4x\sum_q[\bar q(x)\eta_q(x)+\bar\eta_q(x)q(x)]}\times && \\
%e^{i\int d^4x \eta\cdot j(x)\cdot\eta\phi(x)}+O(1/\sqrt{N}g).&&
%\end{eqnarray}
%At this order ghost field just decouples so we can safely ignore it. QCD is so reduced to a Yukawa model by the use of the mapping theorem, at the leading order of a development in the inverse of the 't Hooft coupling. All the parameters of the model are fixed by QCD.
%In the limit $\lambda\rightarrow\infty$, scalar field term takes a Gaussian form [M. Frasca, Phys. Rev. D {\bf 73} (2006) 027701 [hep-th/0511068]]. This theory is trivial in this limit and the beta function goes like $\beta(\lambda)=4\lambda$ in four dimensions at lower momenta.

Now, let us evaluate the two-point function for the Yang-Mills field. One gets immediately
\begin{eqnarray}
\label{eq:Dmunu}
   D_{\mu\nu}^{ab}(x-y)&=&\langle{\cal T}A_\mu^a(x)A_\nu^b(y)\rangle=\eta^a_\mu\eta^b_\nu\langle{\cal T}
   \phi(x)\phi(y)\rangle+O(1/\sqrt{N}g) \nonumber \\
   &=&\eta^a_\mu\eta^b_\nu\Delta(x-y)+O(1/\sqrt{N}g)
\end{eqnarray}
where we have set $\Delta(x-y)=\langle{\cal T}\phi(x)\phi(y)\rangle$ for the two-point function of the scalar field. So, we need to identify the two-point function for the scalar field in the proper limit. Indeed,
we have recently proved \cite{Frasca:2005sx,Frasca:2006yx} that in the limit of the coupling going to infinity, the scalar field reaches a {\em trivial} infrared fixed point and the two-point function is exactly determined as
\begin{equation}
\label{eq:gluonD}
    \Delta(p)=\sum_{n=0}^\infty\frac{B_n}{p^2-m_n^2+i\epsilon}
\end{equation}
being
\begin{equation}
    B_n=(2n+1)\frac{\pi^2}{K^2(i)}\frac{(-1)^{n+1}e^{-(n+\frac{1}{2})\pi}}{1+e^{-(2n+1)\pi}}.
\end{equation}
being $K(i)=\int_0^{\frac{\pi}{2}}\frac{d\theta}{\sqrt{1+\sin^2\theta}}\approx 1.3111028777$, and a formula for the spectrum of the theory, in the strong coupling limit, given by
\begin{equation}
    m_n = \left(n+\frac{1}{2}\right)\frac{\pi}{K(i)}\left(\frac{Ng^2}{2}\right)^{\frac{1}{4}}\Lambda.
\end{equation}
From the mass spectrum we can identify a string tension that will be useful in the following. We set, using the mapping theorem,
\begin{equation}
    \sqrt{\sigma}=\left(\frac{Ng^2}{2}\right)^{\frac{1}{4}}\Lambda=(2\pi N\alpha_s)^{\frac{1}{4}}\Lambda.
\end{equation}
Here $\Lambda$ is an arbitrary parameter arising from the integration of the equations of the theory. So, being an integration constant, it should be obtained from experiment. Finally, we note the following functional expansion for the generating functional of the Yang-Mills theory \cite{Cahill:1985mh,Frasca:2008gi} that holds in a strong coupling limit
\begin{equation}
\label{eq:AD}
    A_\mu^a=\Lambda\int d^4y D_{\mu\nu}^{ab}(x-y)j^{b\nu}(y)+O(j^3).
\end{equation}
The mapping theorem grants that the propagator entering in this equation is the same given in eq.(\ref{eq:gluonD}) with a proper choice of the $\eta$ parameters.
% 28 September 2011 Modified on referee's request 
The form of the propagator shared by the two theories in the infrared limit, provided a 
kind of
K\"allen-Lehman representation
with a non-positive definite spectral function
holds (e.g. see \cite{Strocchi:1993wg}), is an evidence that both theories are infrared trivial. 
This can also be seen by the form of the spectrum that has only free quasi-particle states but no bounded interacting states. This by no means implies that QCD is trivial, rather this theory is infrared safe due to the presence of quarks.

%We just note from this that $\sigma_{SU(2)}/\sigma_{SU(3)}=\sqrt{2/3}$ as seen on lattice\cite{tep1}.
%We recognize that, at lower energies, strong interactions are mediated by a kind of bosons that can be seen as due to Yang-Mills field self-interaction. These are the physical states in a strong coupling limit.

% 29 September 2011 COmpletely new section added

\subsection{Consistency of the choice of the classical solutions}

So far, we have introduced a set of classical solutions and built up on them a quantum field theory without any further support to this choice being the right one or a proof for it. Indeed, we have seen that this construction is self-consistent provided we consider a increasingly large coupling but we cannot claim that other solutions do exist providing a proper description in the same limit or that those gauge configurations grant an optimal saddle point for the path integral of the theory. So, the only way we have to be sure that this picture is the proper one is to compare it with numerical data. On the lattice, very large volumes were considered for the gluon propagator in the Landau gauge in \cite{Bogolubsky:2007ud,Cucchieri:2007md,Oliveira:2007px} while in Ref. \cite{Aguilar:2004sw,Aguilar:2008xm} a numerical solution for Dyson-Schwinger equations was provided.

% 5 November 2011 Added to answer to referee
Our aim will be to show how, increasing the volume, the agreement between numerical data and our analytical results tend to coincide. Numerical Dyson-Schwinger equations represent our infinite volume limit and we expect a very near coincidence of results in this case. 

We consider two kind of lattice computations: A set of volumes till $80^4$ directly obtained with measurements on the lattice for SU(3) and measurements at $128^4$ recovered from figure 2 in \cite{Cucchieri:2007md} for SU(2). We are able to show in this way that, increasing the volume, our propagator describes even more accurately the one measured on the lattice in the deep infrared. We would like to point out that the mass gap is different for these two cases as it depends on the value of $\beta$ that, just for this section, has nothing to do with temperature but is the coupling on the lattice. 
%Then, using numerical Dyson-Schwinger results for SU(3) \cite{Aguilar:2008xm}, where no volume problem arises, we see that our propagator perfectly matches the numerical solution in the deep infrared and deviates from it in the intermediate regime where our approximation is expected to worsen.
% 5 November 2011 Added to answer to referee
We see from the figures below that the situation is the one depicted with volumes to be taken increasingly large on the lattice to match even better our gluon propagator.
Note that we consider a weak dependence on the gauge group as showed in \cite{Maas:2010qw} that is fully consistent with our discussion above.

\begin{figure}[H]
\begin{center}
\includegraphics{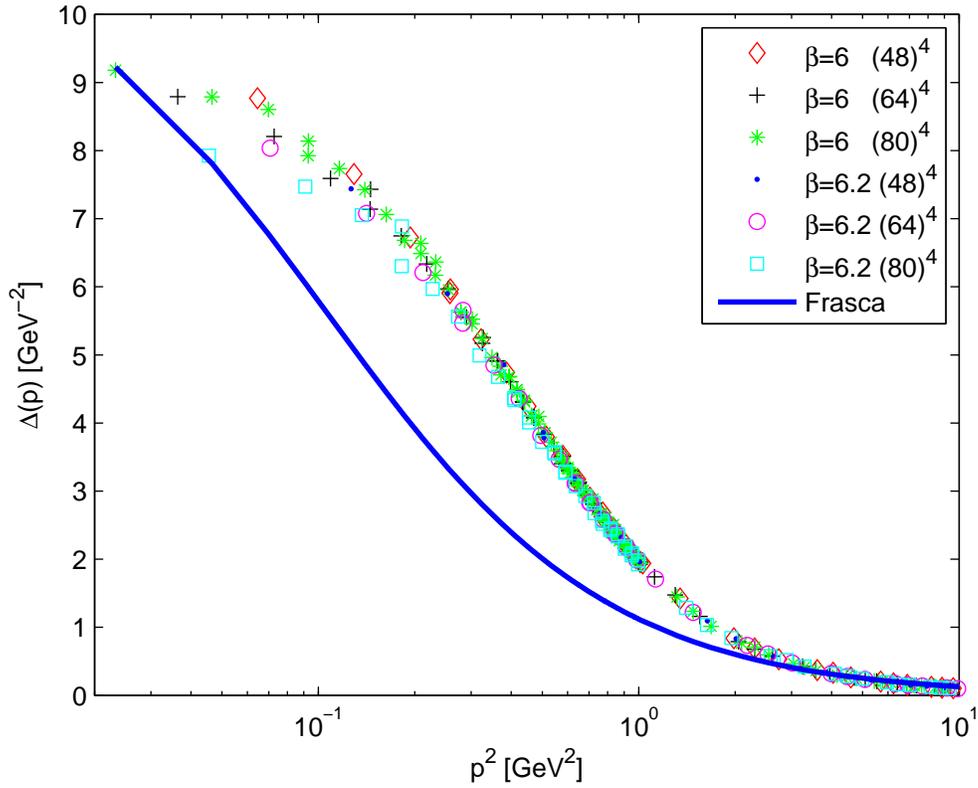}
\caption{(Color online) Gluon propagator in the Landau gauge for SU(3), $\rm 80^4$ with a mass gap of $\rm m_0=321\ MeV$}
\end{center}
\end{figure}

\begin{figure}[H]
\begin{center}
\includegraphics{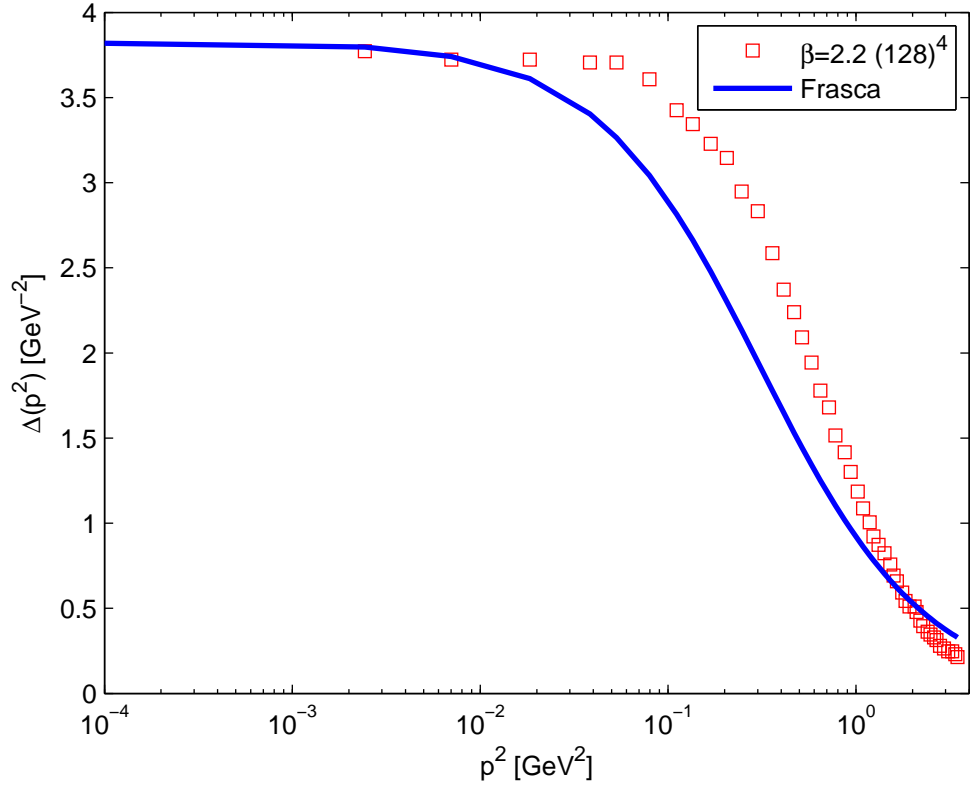}
\caption{(Color online) Gluon propagator in the Landau gauge for SU(2), $\rm 128^4$ with a mass gap of $\rm m_0=555\ MeV$}
\end{center}
\end{figure}

\begin{figure}[H]
\begin{center}
\includegraphics{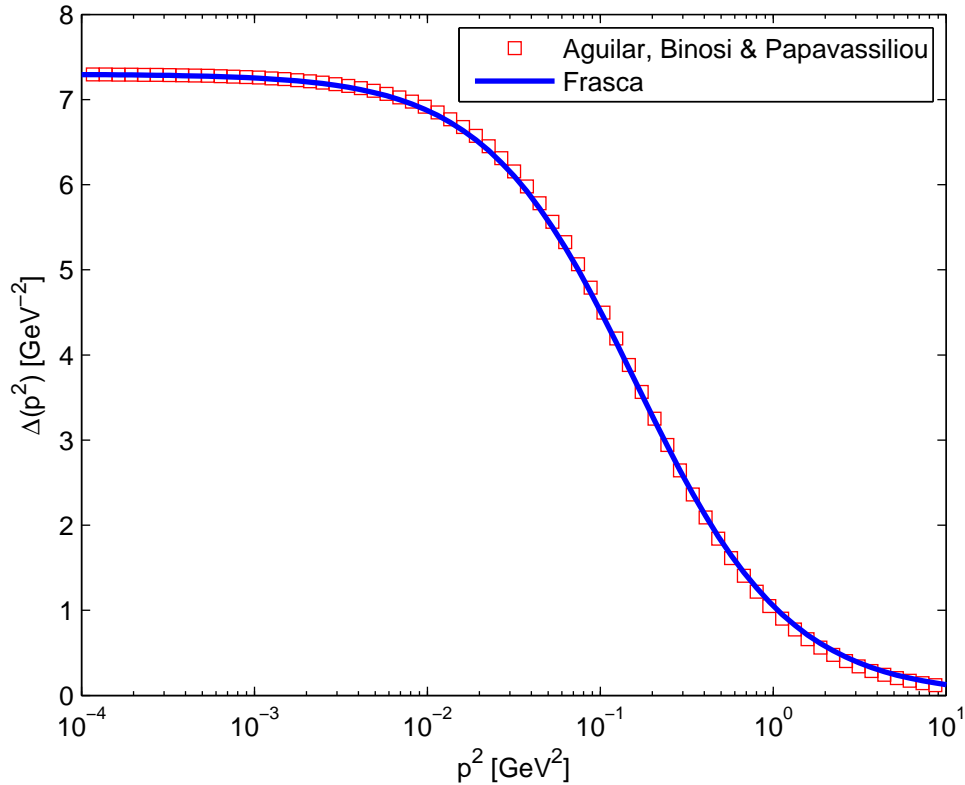}
\caption{(Color online) Gluon propagator in the Landau gauge for SU(3) obtained by numerically solving Dyson-Schwinger equations and a mass gap $\rm m_0=399\ MeV$}
\end{center}
\end{figure}

This agreement between lattice computations at increasing volume and the perfect match for the numerical Dyson-Schwinger equations with our propagator give a strong support to our picture and the view that the classical solutions of our choice provide a correct starting point for a perturbative quantum field theory in the infrared limit.

\subsection{QCD in the infrared limit}

Now, we fix the gauge to Landau through eq.(\ref{eq:etaeta}) and change the potential through eq.(\ref{eq:AD}) obtaining at the leading order, after noting that the ghost field decouples at this order,
\begin{eqnarray}
   S_{QCD}&=&\frac{1}{2}\int d^4xd^4yj^{a\mu}(x)D_{\mu\nu}^{ab}(x-y)j^{b\nu}(y)
   +\int d^4x\sum_q\bar q(x)\left(i\slashed{\partial}\right. \nonumber \\
   &&-g\frac{\lambda^a}{2}\gamma^\mu\Lambda\int d^4y D_{\mu\nu}^{ab}(x-y)j^{b\nu}(y) \nonumber \\
   &&-\left.g^2\Lambda\frac{\lambda^a}{2}\gamma^\mu\int d^4y'D_{\mu\nu}^{ab}(x-y')\sum_{q'}\bar q'(y')\frac{\lambda^b}{2}\gamma^\nu q'(y')\right)q(x) \nonumber \\
   &&+O(1/\sqrt{N}g).
\end{eqnarray}
Now, we use the propagator (\ref{eq:Dmunu}) obtaining in the end
\begin{eqnarray}
   S_{QCD}&=&\frac{1}{2}\int d^4xd^4yj^{a\mu}(x)\Delta(x-y)j^a_\mu(y)
   +\int d^4x\sum_q\bar q(x)\left(i\slashed{\partial}\right. \nonumber \\
   &&\left.-g\frac{\lambda^a}{2}\gamma^\mu\Lambda\int d^4y \Delta(x-y)j^a_\mu(y)\right)q(x) \nonumber \\
   &&-g^2\Lambda\int d^4xd^4y'\Delta(x-y')\sum_q\bar q(x)\frac{\lambda^a}{2}\gamma^\mu q(x)\sum_{q'}\bar q'(y')\frac{\lambda^a}{2}\gamma_\mu q'(y') \nonumber \\
   &&+O(1/\sqrt{N}g).
\end{eqnarray}
So, we see that the existence of the infrared trivial fixed point in a pure Yang-Mills theory has the effect to recover, directly from QCD, a non-local Nambu-Jona-Lasinio model always reducible to a local one \cite{Kondo:2010ts,Frasca:2008zp}. The physics of this model, both at zero and finite temperature, has been fully exploited by Hell, Cristoforetti, Roessner and Weise \cite{Hell:2008cc} with a substantial difference that they are forced to guess the form of the form factor (the gluon propagator) using a model of a liquid of instantons. Here the form factor is directly obtained from QCD but we will see below that their guess is excellently good.

\section{Non-local Nambu-Jona-Lasinio model and gap equation}
\label{sec2}

\begin{figure}[H]
\begin{center}
\includegraphics{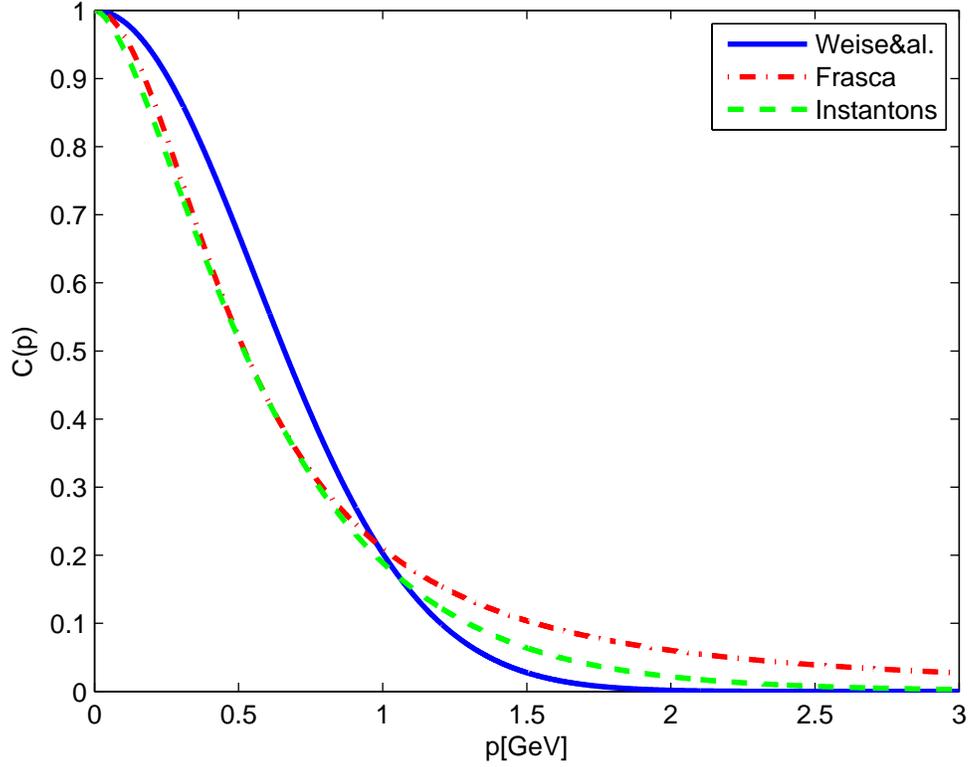}
\caption{\label{fig:weise} (Color online) Comparison between our form factor with that used in Ref.\cite{Hell:2008cc} and the instanton liquid in Ref.\cite{Schafer:1996wv} used as a model in Ref.\cite{Hell:2008cc}.}
\end{center}
\end{figure}

In order to define completely the model we have to try to analyze better the propagator. We realize from eq.(\ref{eq:gluonD}) that higher excited states are exponentially damped and so we can limit our analysis to a single scalar field interacting with quarks. So, we approximate the propagator as $\Delta(p)\approx B_0/(p^2-m_0^2+i\epsilon)$, being $m_0\approx 1.19\sqrt{\sigma}$ and $\sigma=(0.44\ GeV)^2$ the string tension, and neglect the other contributions coming from higher excited states. This means that the Gaussian term $\frac{1}{2}\int d^4xd^4yj^{a\mu}(x)\Delta(x-y)j^a_\mu(y)$ can be rewritten, in this approximation using an arbitrary scalar field $\sigma$ that we integrate over, as $\frac{1}{2}\int d^4x\left[(\partial\sigma)^2-m_0^2\sigma^2\right]$, provided we take
\begin{equation}
   \sigma=\sqrt{3(N^2-1)/B_0}\Lambda\int d^4y \Delta(x-y)j(y)
\end{equation}
and writing down the currents as $j_\mu^a=\eta_\mu^a j$. So, one finally has
\begin{eqnarray}
   S_{QCD}&=&\frac{1}{2}\int d^4x\left[\frac{1}{2}(\partial\sigma)-\frac{1}{2}m_0^2\sigma^2\right]
   +\int d^4x\sum_q\bar q(x)\left(i\slashed{\partial}\right. \nonumber \\
   &&\left.-g\sqrt{\frac{B_0}{3(N^2-1)}}\frac{\lambda^a}{2}\gamma^\mu\eta_\mu^a\sigma(x)\right)q(x) \nonumber \\
   &&-g^2\Lambda\int d^4xd^4y'\Delta(x-y')\sum_q\bar q(x)\frac{\lambda^a}{2}\gamma^\mu q(x)
   \sum_{q'}\bar q'(y')\frac{\lambda^a}{2}\gamma_\mu q'(y') \nonumber \\
   &&+O(1/\sqrt{N}g).
\end{eqnarray}
We get a coupling for the $\sigma$ field that can be ignored for our aims. In order to recover in full the non-local model of ref.\cite{Hell:2008cc} we have to identify the form factor depending on the gluon propagator. We get immediately
\begin{equation}
   {\cal G}(p)=-\frac{1}{2}g^2\sum_{n=0}^\infty\frac{B_n}{p^2-(2n+1)^2(\pi/2K(i))^2\sigma+i\epsilon}=\frac{G}{2}{\cal C}(p)
\end{equation}
being $G$ the Nambu-Jona-Lasinio constant that in our case is given by $G=2{\cal G}(0)=(g^2/\sigma)\sum_{n=0}^\infty\frac{B_n}{(2n+1)^2(\pi/2K(i))^2}\approx 0.7854(g^2/\sigma)$, so that ${\cal C}(0)=1$, definitely fixed by QCD. 
% 28 September 2011 extended and modified the figure on request of the referee
In Ref.\cite{Hell:2008cc}, a guess was put forward for ${\cal C}(p)$ using a model of liquid of instantons. 
In Ref.\cite{Schafer:1996wv} the form factor for this case takes the form
\begin{equation}
\mathcal{C}_I(p)=p^2\left\{\pi d^2 \dfrac{d}{d\xi}\big[I_0(\xi)K_0(\xi)-I_1(\xi)K_1(\xi)\big]\right\}^2\qquad\text{with } \xi=\frac{|p| d}{2} 
\end{equation}
being $I_n$ and $K_n$ Bessel functions. In the following we normalize this function to be 1 at zero momenta dividing it by ${\cal C}_I(0)$.
%
%They
%
Weise et al.
fix the functional form to ${\cal C}(p)=\exp(-p^2d^2/2)$ with $d^{-1}\approx 0.56\ GeV$
in order to avoid too much computational weight
. We compared our ${\cal C}(p)$ with that given in Ref.\cite{Hell:2008cc}, fixing $\sigma=(0.44\ GeV)^2$
for the string tension
%
%, and
and rather taking $d^{-1}\approx 0.58\ GeV$, not much different from Weise et al. guess.
The result is presented in fig.\ref{fig:weise}. The agreement is 
%so good indeed 
so strikingly good with the instanton form factor
that our conclusions strongly support a description of the ground state of QCD as an instanton liquid. This result was already pointed out in \cite{Frasca:2008gi} by comparison with lattice results \cite{Boucaud:2002fx}
for the running coupling in the infrared limit
.

With the given expression for the form factor, representing one of the most important results given in this paper, we are able to put down the gap equation for massless quarks as obtained from ref.\cite{Hell:2008cc}:
\begin{equation}
    M(p)={\cal C}(p)v
\end{equation}
and
\begin{equation}
    v=\frac{4NN_f}{m_0^2+1/G}\int\frac{d^4p}{(2\pi)^4}{\cal C}(p)\frac{M(p)}
    {p^2+M^2(p)}
\end{equation}
being $v$ the v.e.v. of the $\sigma$ field, $N$ the number of colors and $N_f$ the number of flavors. Our aim is to prove the existence of a phase transition at finite temperature.

\section{Finite temperature gap equation}
\label{sec3}

We can evaluate the above results at finite temperature by passing to Matsubara sums. So, finally we can write down the gap equation as
\begin{equation}
    M(\omega_k,{\bm p})={\cal C}(\omega_k,{\bm p})v
\end{equation}
and
\begin{equation}
\label{eq:v0}
    v=\frac{4NN_f}{m_0^2+1/G}\beta^{-1}\sum_{k=-\infty}^\infty\int\frac{d^3p}{(2\pi)^3}{\cal C}(\omega_k,{\bm p})\frac{M(\omega_k,{\bm p})}
    {\omega_k^2+{\bm p}^2+M^2(\omega_k,{\bm p})}
\end{equation}
being the Matsubara frequencies $\omega_k=(2n+1)\pi T$ with $n$ an integer. The limits we are interested in are those at small momenta and temperature. The first one is needed for consistency with the Nambu-Jona-Lasinio model while the second is needed to identify the existence of a phase transition. Now, we are in a position to prove the existence of a critical point for which $v=0$ and the chiral symmetry is restored. Setting $v=0$ into eq.(\ref{eq:v0}), we have to solve
\begin{equation}
\label{eq:v}
    \frac{4NN_f}{m_0^2+1/G}\beta^{-1}\sum_{k=-\infty}^\infty\int\frac{d^3p}{(2\pi)^3}{\cal C}^2(\omega_k,{\bm p})\frac{1}
    {\omega_k^2+{\bm p}^2}=1.
\end{equation}
A possible study of this equation is through numerical techniques. But taking a look at fig.\ref{fig:weise}, after a simple numerical evaluation, we note that the form factor is about 0.8 for a momentum of 260 GeV. Lattice computations as those in \cite{Fodor:2001pe,Aoki:2009sc} estimate the critical temperature at about 170 MeV, well below the limit where the form factor is approximated by unity. This means that, for our aims, the form factor can be reduced to a step function cutting at zero at about 300 MeV and being unity for lower energies. With this crude approximation we are able to get an analytic expression for the critical temperature. Indeed, in this case the integral can be exactly evaluated and we get
\begin{equation}
    T_c^2\approx\frac{3}{\pi^2}\left[\Lambda^2-\frac{\pi^2}{NN_f}\left(m_0^2+\frac{1}{G}\right)\right]
\end{equation}
that proves, starting directly from QCD, that a critical point indeed exists for which chiral symmetry is broken. This formula is in close agreement with the one in a recent work by Scoccola and G\'omez Dumm \cite{GomezDumm:2004sr}. The main difference is that we have all fixed through the proper value of the mass gap $m_0$ due to the form factor.

Now, we can get an estimation of $\Lambda$, a parameter otherwise fixed by experiment for the Nambu-Jona-Lasinio model, just fixing $T_c = 0.17\ GeV$ given by lattice computations. Taking $\sigma = (0.44\ GeV)^2$, $g\approx 3$, $N=3$ and $N_f=2$ we get $\Lambda=0.77\ GeV$, a perfectly reasonable value for the Nambu-Jona-Lasinio model. This value decreases by increasing the number of flavors.
%Similarly, for $T_c = 0.15\ GeV$ we get $\Lambda=0.567\ GeV$. 
So, from this computation we can conclude that both groups in \cite{Aoki:2009sc,Bazavov:2009zn} get a perfectly reasonable value for the cross-over temperature, providing this can be maintained at zero quark masses and chemical potential.

\section{Conclusions}
\label{sec4}
%%%%%%%%%%%%%%%%
%%%%%%%%%%%%%%%%%%%%%%%%%%%

We have successfully shown how, starting from the quantum field theory of QCD, this theory reduces, in the low-energy limit, to a non-local Nambu-Jona-Lasinio model having all the parameters properly fixed to physical values. Besides, the form of the gluon propagator in such a low-energy limit is exactly known. This has implied that, when the computation is extended to a finite temperature case, a proof of existence of a phase transition, with chemical potential and quark masses set to zero, is given.

This result should only be considered a starting point for future analysis. The most important of this is that we have not been able yet to accomodate the Polyakov loop in this approach. Presently, this is an important tool for the understanding of the phase diagram of QCD. Besides, it will be interesting to see how the cross-over emerges assuming a quark mass different from zero and making the gap equation more involved just introducing a chemical potential. This is our working program for the near future.

\section*{Acknowledgements}
A lot of discussions about this matter have been made with Marco Ruggieri. I have to thank him very much for clarifying several points in this matter that otherwise would have remained obscure.
I would like to thank Thomas Hell for pointing me out the right equation to use for instantons liquid that is also the one that was used in his paper with Weise, Roessner and Cristoforetti. 
Finally, I would like to thank Orlando Oliveira for providing me his measurements of the gluon propagator on the lattice, Arlene Aguilar and Daniele Binosi for providing me the numerical results of their work on Dyson-Schwinger equations.

%\newpage

% THANK YOU VERY MUCH CHARLES!!!
%\begin{acknowledgments}
%\end{acknowledgments}

%\bibliography{pra001}

\end{document}